\documentclass[seceq]{ptptex}

\usepackage{graphicx}

\usepackage{wrapft}

\title{
Coupled Oscillator Model of the Business Cycle with Fluctuating Goods Markets
}

\author{
Y.~\textsc{Ikeda}$^{1}$,
H.~\textsc{Aoyama}$^{2}$,
Y.~\textsc{Fujiwara}$^{3}$,
H.~\textsc{Iyetomi}$^{4}$,
K.~\textsc{Ogimoto}$^{1}$,
W.~\textsc{Souma}$^{5}$,
and H.~\textsc{Yoshikawa}$^{6}$
}

\inst{
$^1$Institute of Industrial Science, University of Tokyo, Tokyo 153-8505, Japan\\
$^2$Graduate School of Sciences, Kyoto University, Kyoto 606-8502, Japan\\
$^3$Graduate School of Simulation Studies, University of Hyogo, Kobe 650-0047, Japan\\
$^4$Department of Physics, Niigata University, Niigata 950-2181, Japan\\
$^5$College of Science and Technology, Nihon University, Chiba 274-8501, Japan\\
$^6$Faculty of Economics, University of Tokyo, Tokyo 113-0033, Japan\\
}

\recdate{
}

\abst{
The sectoral synchronization observed for the Japanese business cycle in the Indices of Industrial Production data is an example of synchronization.
The stability of this synchronization under a shock, e.g., fluctuation of supply or demand, is a matter of interest in physics and economics.
We consider an economic system made up of industry sectors and goods markets in order to analyze the sectoral synchronization observed for the Japanese business cycle.
A coupled oscillator model that exhibits synchronization is developed based on the Kuramoto model with inertia by adding goods markets, and
analytic solutions of the stationary state and the coupling strength are obtained.
We simulate the effects on synchronization of a sectoral shock for systems with different price elasticities and the coupling strengths.
Synchronization is reproduced as an equilibrium solution in a nearest neighbor graph. 
Analysis of the order parameters shows that the synchronization is stable for a finite elasticity,
whereas the synchronization is broken and the oscillators behave like a giant oscillator with a certain frequency additional to the common frequency for zero elasticity. 
}

\begin{document}
\maketitle

\section{Introduction}

The business cycle is an example of synchronization, and this has been studied in nonlinear physics. In particular, the stability of synchronization under a shock, e.g., a fluctuation in supply, is a matter of interest.
Sectoral synchronization was observed for the Japanese business cycle in the Indices of Industrial Production data from 1988 to 2007. \cite{Iyetomi2011a} \cite{Iyetomi2011b} 
By synchronization, we means there is a constant phase difference between industry sectors. 
For instance, if random noise components are removed in an appropriate manner, the business cycle is written as 
\begin{equation}
x_i = \sin (\omega t + \theta_i),
\label{eq:Grate}
\end{equation}
\begin{equation}
\omega = \frac{2 \pi}{T},
\label{eq:Omega}
\end{equation}
where $x_i$, $\theta_i$, $\omega$, and $T$ are the normalized growth rate of production for sector i, the phase of the business cycle for sector i, the common angular frequency, and the common period of the business cycle, respectively. 

We consider an economic system made up of industry sectors and goods markets. Various types of shock may occur in the economic system
of interest. If production in a specific sector suddenly decreases significantly, the imbalance of demand and supply may destroy the synchronization. 
However, if the price of the goods in the market is quickly adjusted, a sudden change of production in the sector will be absorbed by consumers as a decrease in demand in coupled sectors. 

In this paper, 
we describe a synchronized coupled oscillator model with flucuating goods markets. 
The effects on synchronization with respect to the magnitude of shock, price flexibility, and network topology are studied using the coupled oscillator model.
The paper is organized as follows. 
In section 2, existing theories are briefly reviewed.
In section 3, we explain the formulation of our coupled oscillator model,
and we describe our analysis of sectoral synchronization in section 4.
Finally, section 5 presents our conclusions.

\section{Existing Theories}

In this section, existing theories for the synchronization of economic systems, power systems, and physical and biological systems are reviewed. It is noted that the first two are large-scale manmade systems that exhibit similar dynamical behavior in terms of synchronization. 

\subsection{Economic Systems}

The business cycle is observed in most of industrialized economies. Economists have studied this phenomenon by means of mathematical models, including various kinds of linear non-linear, and coupled oscillator models.

Interdependence, or coupling, between industries in the business cycle has been studied for more than half a century.
A study of the linkages between markets and industries using nonlinear difference equations suggests a dynamical coupling among industries. \cite{Goodwin1947}
A nonlinear oscillator model of the business cycle was then developed using a nonlinear accelerator as the generation mechanism. \cite{Goodwin1951}
In this paper, we stress the necessity of nonlinearity because linear models are unable to reproduce sustained cyclical behavior, and tend to either die out or diverge to infinity. 

However, it is noted that a simple linear economic model, based on ordinary economic principles, optimization behavior, and rational expectations, can produce cyclical behavior much like that found in business cycles. \cite{LongPlosser1983} 
An important question aside from synchronization in the business cycle is whether sectoral or aggregate shocks are responsible for the observed cycle. This question has been  empirically examined, and it was clarified that business cycle fluctuations are caused by small sectoral shocks, rather than by large common shocks. \cite{LongPlosser1987}

As the third category of model, coupled oscillators were developed in order to study noisy oscillating processes like national economies. \cite{Anderson1999} \cite{Selover2003}
Simulations and empirical analyses showed that synchronization between the business cycles of different countries is consistent with such mode-locking behavior. 
Along this line of approach, a nonlinear mode-locking mechanism was further studied that described a synchronized business cycle between different industrial sectors. \cite{Sussmuth2003}

\subsection{Power Systems}

Power systems made up of many synchronous machines and transmission lines exhibits synchronization.
The stability and control of this synchronization of the machines has been studied in power system engineering. \cite{Kundur1993}

\begin{figure}
\begin{center}
\includegraphics[width=0.45\textwidth]{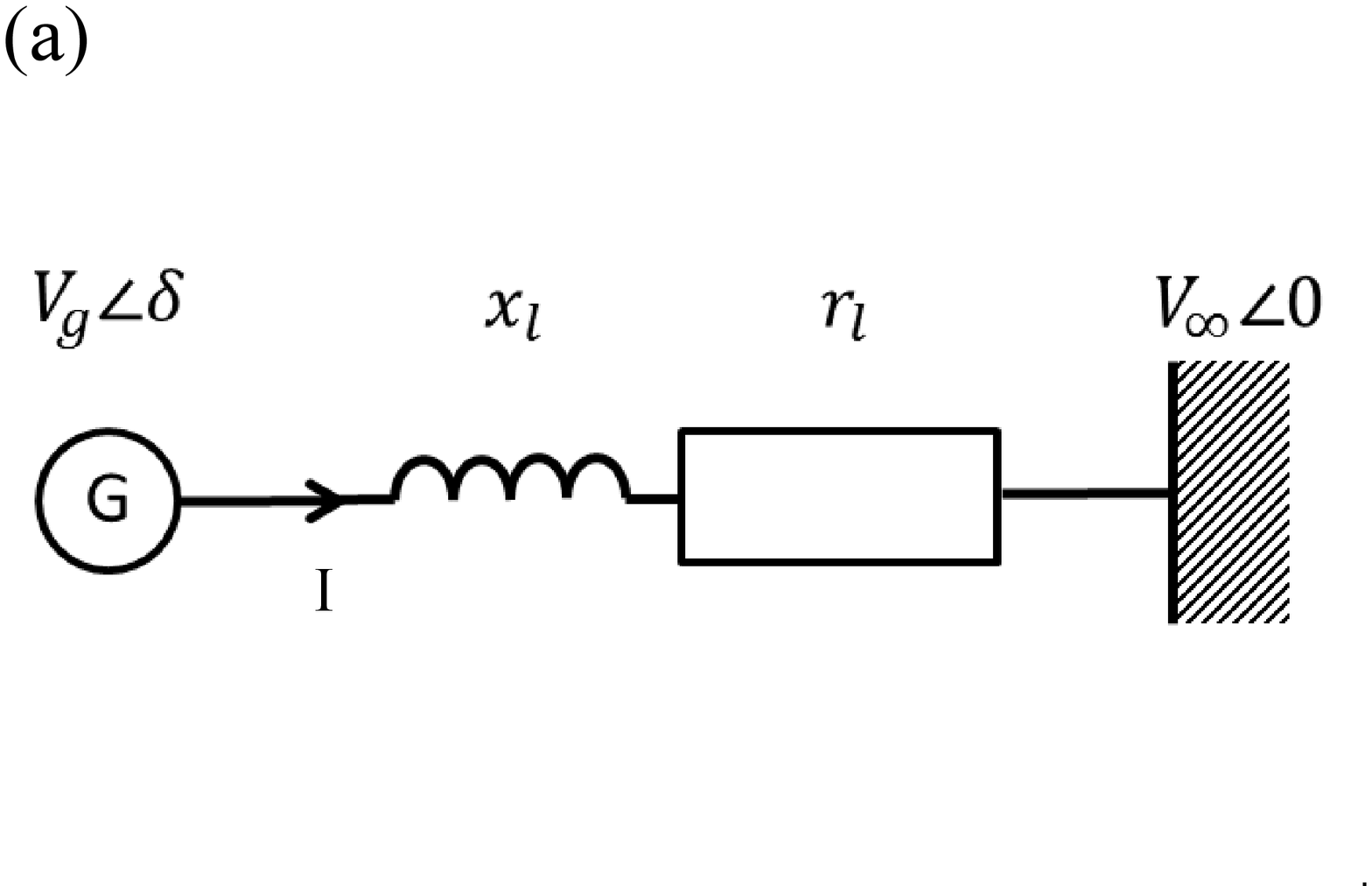}
\includegraphics[width=0.45\textwidth]{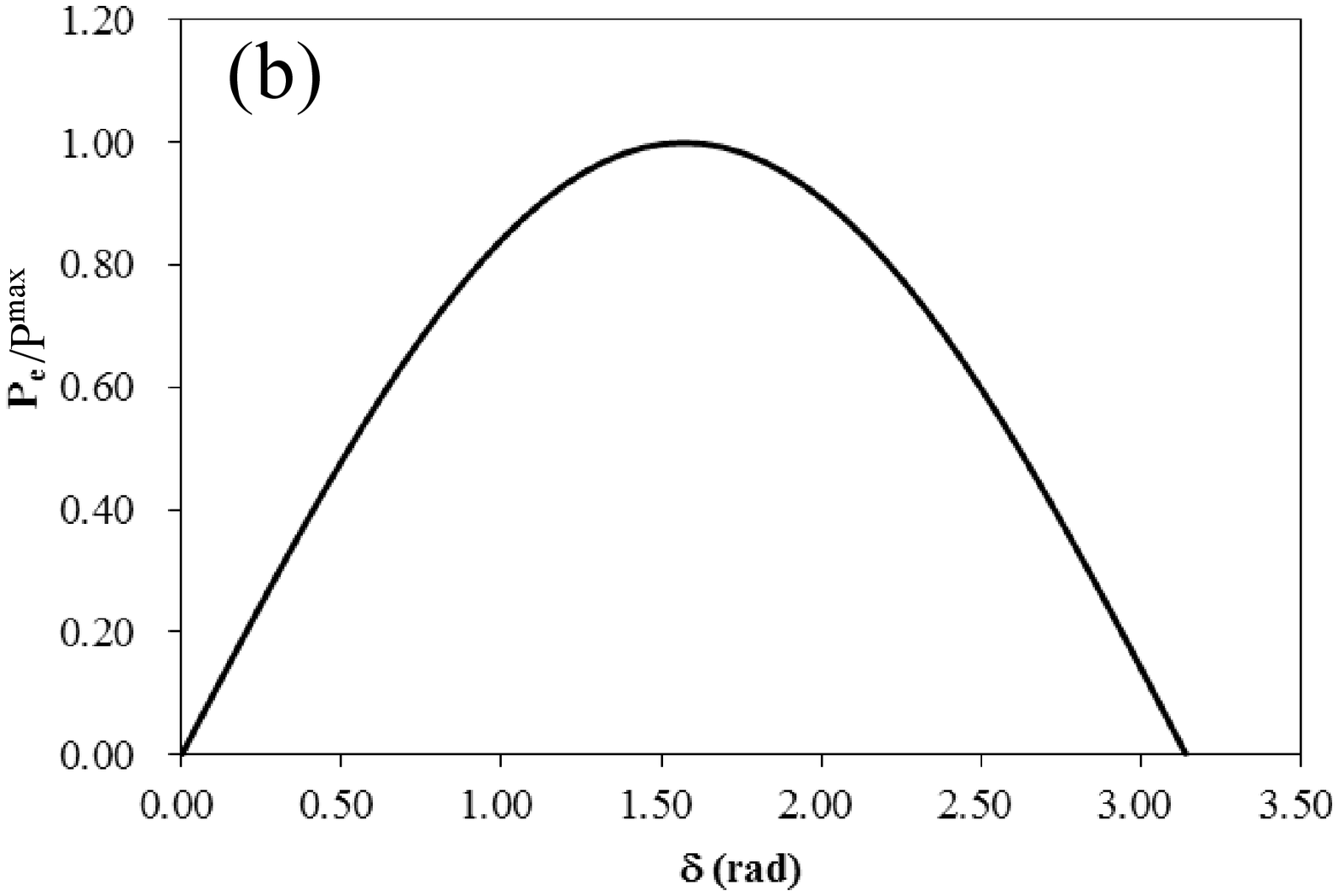}
\caption{Single Machine connected to Infinite Bus and Synchronizing Force}
\label{fig:Synchro}
\end{center}
\end{figure}

The dynamical property of a single  synchronous machine is often studied in the single machine connected to the infinite bus shown in Fig. \ref{fig:Synchro} (a), 
where $V_g$, $\angle \delta$, $I$, $x_l$, $r_l$, $V_\infty$, and $\angle 0$ are the generator voltage, generator phase, current, transmission line reactance, transmission line resistance, infinite bus voltage, and infinite bus phase, respectively.
Here the infinite bus is characterized by constant voltage and constant frequency.
The active power $P_e$ and reactive power $Q_e$ are calculated by 
\begin{equation} 
P_e + i Q_e= V_g exp (i \delta) I^*, 
\label{eq:PQ} 
\end{equation}%
where $*$ indicates the complex conjugate, and the current $I$ is given  by 
\begin{equation} 
I = \frac{V_g exp (i \delta) - V_\infty}{r_l + i x_l}.
\label{eq:I} 
\end{equation}%
By substituting Eq. (\ref{eq:I}) into Eq. (\ref{eq:PQ}), the active power $P_e$ is obtained as the real part of the equation
\begin{equation} 
P_e = \frac{V_g V_{\infty}}{x_l} \sin \delta = P^{max} \sin \delta,
\label{eq:Pe} 
\end{equation}%
by considering relation $r_l \ll x_l$.
$P_e$ is plotted as a function of the generator phase $\delta$ in Fig. \ref{fig:Synchro} (b).
It should be noted that $P_e$ is the electric power transmitted through the transmission line and works as a synchronizing force in power systems.

\subsection{Physical and Biological Systems}

Many collective synchronization  phenomena are known in physical and biological systems. \cite{Strogatz2000}
Physical examples include clocks hanging on a wall, an array of lasers, microwave oscillators, and Josephson junctions.
Biological examples include synchronously flashing fireflies, networks of pacemaker cells in the heart, and metabolic synchrony in yeast cell suspensions.

Kuramoto proposed a coupled oscillator model to explain this rich variety of synchronization phenomena. \cite{Kuramoto1975} \cite{Strogatz2000} \cite{Acebron2005} 
In his model, the dynamics of the oscillators are governed by
\begin{equation}
\dot{\theta_i} = \omega_i + \sum_{j=1}^{N}  k_{ji} \sin(\theta_j - \theta_i),
\label{eq:Kuramoto}
\end{equation}
where $\theta_i$, $\omega_i$, and $k_{ji}$ are the oscillator phase, the natural frequency, and the coupling strength, respectively.
The second term of the RHS of Eq. (\ref{eq:Kuramoto}) is identical to Eq. (\ref{eq:Pe})
If the coupling strength $k_{ij}$ exceeds a certain threshold, the system exhibits synchronization.

\section{Coupled Oscillator Model}

In this section, we first explain the formulation of our coupled oscillator model, and then give an illustrative example to understand the basic behavior of the model.

\subsection{Formulation}

Our model is developed based on the Kuramoto model with inertia \cite{Filatrella2008} by adding goods markets.
We consider the system to consist of oscillator $i$ and oscillator $j$.
The angle $\theta_i$ of oscillator $i$ is written as
\begin{equation} 
\theta_i = \omega t + \tilde{\theta}_i,
\label{eq:CO1} 
\end{equation}%
where $\omega$ and $\tilde{\theta}_i$ are the frequency and phase, respectively. 

The energy dissipated as heat from oscillator $i$ at a rate proportional to the square of the angular velocity is 
\begin{equation} 
P_{d} = K_D (\dot{\theta}_i)^2, 
\label{eq:CO2} 
\end{equation}%
and the kinetic energy accumulated in oscillator $i$ at a rate proportional to the square of the angular velocity is 
\begin{equation} 
P_{a} = \frac{1}{2} I \frac{d}{dt}(\dot{\theta}_i)^2, 
\label{eq:CO3} 
\end{equation}%
where $K_D$ and $I$ are a dissipation constant and a moment of inertia, respectively.
From Eq. (\ref{eq:CO1}), the angular difference $\Delta \theta_{ji}$ is written as the phase difference $\Delta \theta_{ji} = \theta_j - \theta_i = \tilde{\theta}_j - \tilde{\theta}_i$ using the phase $\tilde{\theta}_i$. 
The power transmitted from one oscillator to another is given by
\begin{equation} 
P_{t} = -k_{ji} \sin \Delta \theta_{ji}, 
\label{eq:CO5} 
\end{equation}%
using the phase difference $\Delta \theta_{ji}$. 
In Eq. (\ref{eq:CO5}), the negative sign is used to indicate that power is lost from oscillator $i$.

By substituting relations (\ref{eq:CO2}), (\ref{eq:CO3}), and (\ref{eq:CO5}) into the power balance equation for oscillator $i$ ($P_{s} = P_{d} + P_{a} + P_{t}$), 
we obtain an equation corresponding to Eq. (\ref{eq:Kuramoto}) 
\begin{equation} 
I \omega \ddot{\tilde{\theta}}_i = P_{s} - K_D \omega^2  - 2 K_D \omega \dot{\tilde{\theta}}_i + k_{ji} \sin \Delta \theta_{ji}, 
\label{eq:CO10} 
\end{equation}
using the approximate relation $\dot{\tilde{\theta}}_i \ll \omega$.
From the above discussion, without loss of generality, we obtain an equation describing the dynamics of an $N$-oscillator system,
\begin{equation} 
\ddot{\tilde{\theta}}_i = P_i - \alpha \dot{\tilde{\theta}}_i + \sum_{j=1}^N k_{ji} \sin \Delta \theta_{ji},
\label{eq:CO11} 
\end{equation}
where $I \omega$, $P_{s} - K_D \omega^2$, and $2 K_D \omega$ are replaced by $1$, $P_i$, and $\alpha$, respectively.
$P_i$ is regarded as the net input to oscillator $i$, which is equal to the difference between the input to oscillator $i$ from outside the $N$-oscillator system and the output from oscillator $i$ to outside the $N$-oscillator system.
Hereafter $\tilde{\theta}_i$ is written as $\theta_i$ for simplicity.

\begin{figure}
\begin{center}
\includegraphics[width=0.45\textwidth]{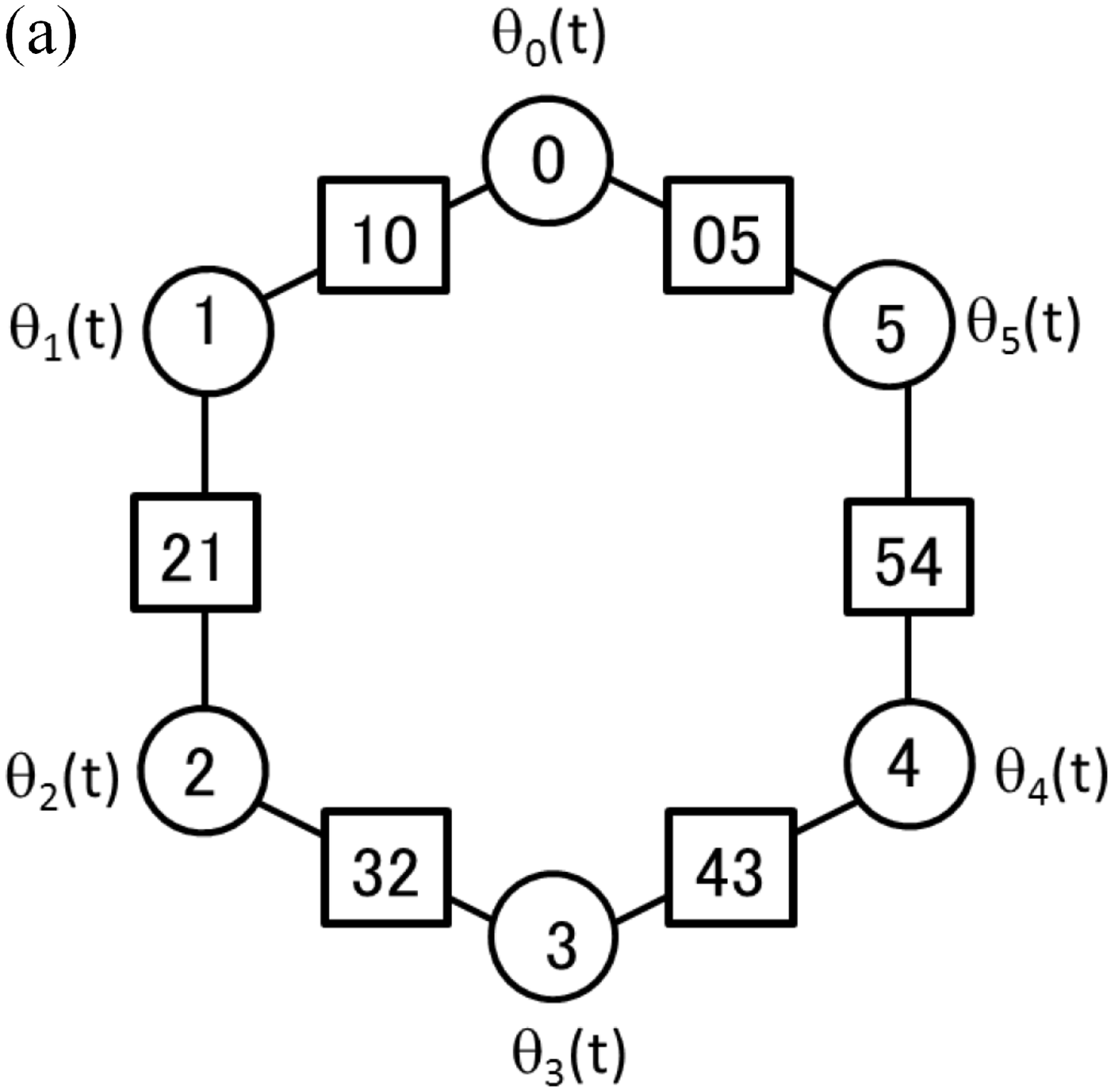}
\includegraphics[width=0.45\textwidth]{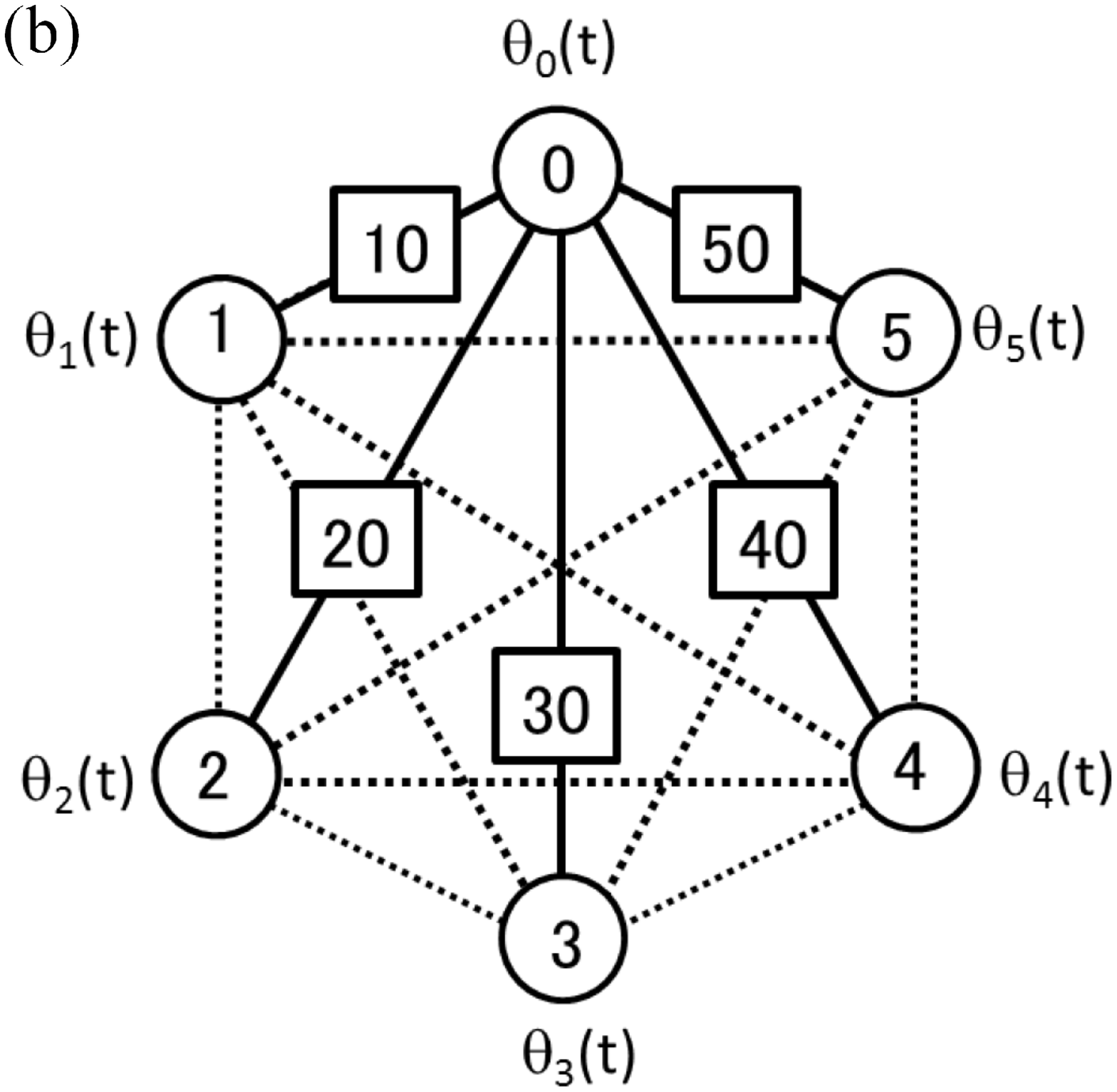}
\caption{Network Topology}
\label{fig:Network}
\end{center}
\end{figure}

It is noted that the synchronizing force is interpreted as being the goods markets. 
The meaning of the goods markets is clarified by adding the sectoral fluctuations of demand or supply, $\delta_{ji} $, as 
\begin{equation}
\ddot{\theta_i} = P_i - \alpha \dot{\theta_i} +  \sum_{j=1}^{N} \{ k_{ji} \sin \Delta \theta_{ji} + \delta_{ji} \}.
\label{eq:OscillatorN}
\end{equation}
The introduction of these sectoral fluctuations is intended to be consistent with the importance of the small sectoral shocks found in Ref. \cite{LongPlosser1987}.
Figure \ref{fig:Network} depicts a system consisting of six oscillators. 
Figure \ref{fig:Network} (a) is the Nearest Neighbor (NN) graph, and Fig. \ref{fig:Network} (b) is the Complete (C) graph.
Oscillators and goods markets are indicated by circles and rectangles, respectively.
In Fig. \ref{fig:Network} (a), the number of the markets is six, because they are only open to the neighboring oscillators.
In Fig. \ref{fig:Network} (b), however, the number of markets is five for each oscillator and the total number of markets is equal to $(6 \times 5)/2=15$. 

Demand $d_{ij}$ for, and supply $s_{ji}$ of, good $i$ can be written as
\begin{equation}
d_{ij} = d_0 + \delta_{ij}, 
\label{eq:dij}
\end{equation}
\begin{equation}
s_{ji} = s_0 + \delta_{ji}, 
\label{eq:sji}
\end{equation}
and are determined through the goods market $ij$.
$d_0$ and $s_0$ are the equilibrated demand and equilibrated supply of good $i$ in goods market $ji$, 
and $\delta_{ij}$ and $\delta_{ji}$ are the sectoral fluctuations of demand and supply, respectively. 
In Eqs. (\ref{eq:dij}) and (\ref{eq:sji}) the equilibrated demand $d_0$ and equilibrated supply $s_0$ are given by
\begin{equation}
d_0 = s_0 = k_{ji} | \sin( \theta_j^* - \theta_i^* ) |,
\label{eq:d0}
\end{equation}
where the equilibrated phase $\theta_i^* (i=1, \cdots, N)$ are solution of Eq. (\ref{eq:OscillatorN}) with $\ddot{\theta_i}=\dot{\theta_i}=0$.
The relations between demand $d$ or supply $s$ and the price $p$ of good $i$ are written 
using the price elasticity of demand $\epsilon_d$ or the price elasticity of supply $\epsilon_s$, 
\begin{equation}
\frac{d}{d_0} = \left( \frac{p}{p_0} \right)^{\epsilon_d},
\label{eq:DemandElastisity}
\end{equation}
\begin{equation}
\frac{s}{s_0} = \left( \frac{p}{p_0} \right)^{\epsilon_s}.
\label{eq:SupplyElastisity}
\end{equation}

\begin{figure}
\begin{center}
\includegraphics[width=0.45\textwidth]{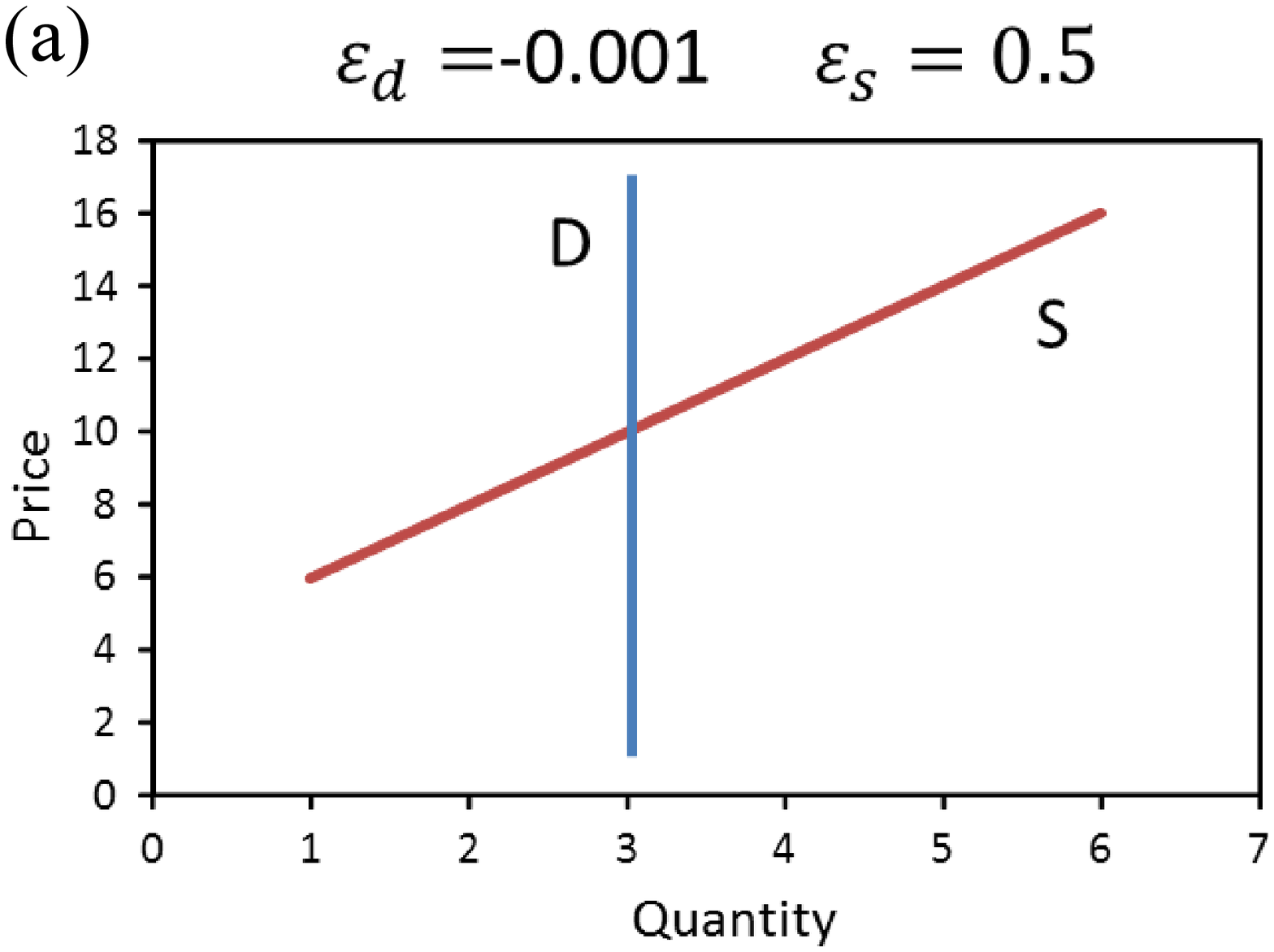}
\includegraphics[width=0.45\textwidth]{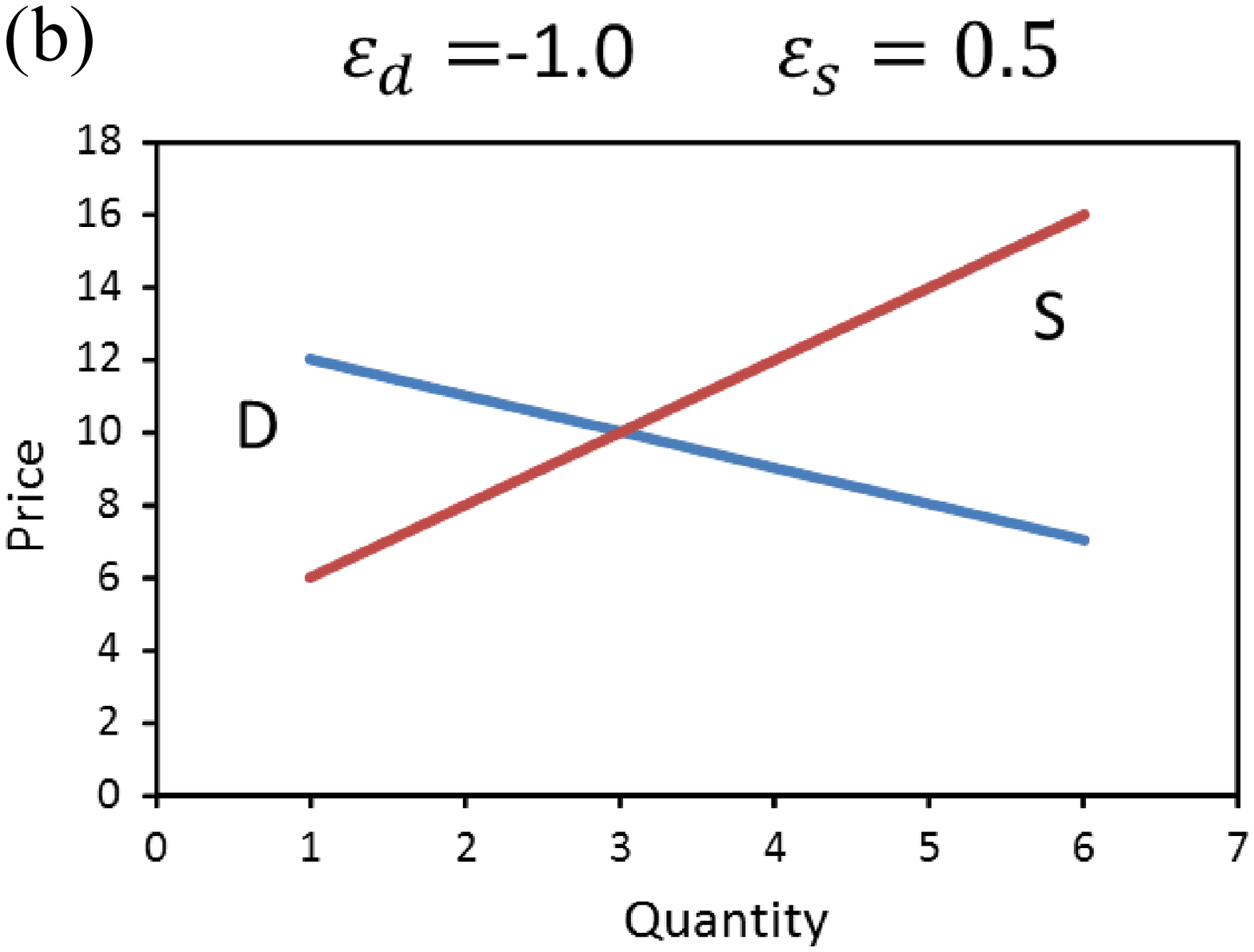}
\caption{Demand and Supply}
\label{fig:Demand and Supply}
\end{center}
\end{figure}

The price as a function of demand and supply is depicted in Fig. \ref{fig:Demand and Supply}. 
Figure \ref{fig:Demand and Supply} (a) is plotted for small value of $\epsilon_d$. In this case, the demand is shown as a vertical line.
Oscillator $j$ responds to the sectoral fluctuations of supply $\delta_{ji}$ according to price $p$.
Depending on the market flexibility, oscillator $j$ changes its demand by $\delta_{ij}$, given by
\begin{equation}
\delta_{ij} =\begin {cases}
				- \delta{ji} & (\epsilon_d < 0) \\
				0 & (\epsilon_d = 0),
				\end{cases}
\label{eq:deltaij}
\end{equation}
by responding to the fluctuation of supply $\delta_{ji}$. 

\subsection{Illustrative example}

We consider an oscillator system in order to understand the basic behavior of the coupled oscillator model.
For the NN graph, the model equations are written as 
\begin{align}
\ddot{\theta}_1 &= P_1 - \alpha \dot{\theta}_1 + \{ k_{21} \sin(\theta_2 - \theta_1) + \delta_{21} \}, \
\notag \\
\ddot{\theta}_2 &= P_2 - \alpha \dot{\theta}_2 + \{ k_{21} \sin(\theta_1 - \theta_2) + \delta_{12} \} + \{ k_{32} \sin(\theta_3 - \theta_2) + \delta_{32} \}, \
\notag \\
& \cdots \label{eq:Oscillator}  \\
\ddot{\theta}_{N-1} &= P_{N-1} - \alpha \dot{\theta}_{N-1} + \{ k_{N-1 N-2} \sin(\theta_{N-2} - \theta_{N-1}) + \delta_{N-2 N-1} \}, \
\notag \\
&+ \{ k_{N N-1} \sin(\theta_N - \theta_{N-1}) + \delta_{N N-1} \}, \
\notag \\
\ddot{\theta}_N &= P_N - \alpha \dot{\theta}_N + \{ k_{N N-1} \sin(\theta_{N-1} - \theta_N) + \delta_{N N-1} \}.
\notag
\end{align}
Analytic solutions of the stationary state will be obtained by solving the simultaneous equations of (\ref{eq:Oscillator}) with $\dot{\theta}_i=\ddot{\theta}_i=0 (i=1, \cdots, N)$.
We consider the case, where $P_1=P=1$, $P_i=0 (i=2, \cdots, N-1)$, $P_N=-P=-1$, and $\theta_N=0$. 
The analytic solutions obtained for the stationary state $\theta_i^*$ and the synchronizing coupling strengths $k_{ji}$ are
\begin{equation}
\theta_i^* =  \theta_{i+1}^* + \arcsin \frac{P}{k_{ji}},
\label{eq:StatSol}
\end{equation}
\begin{equation}
k_{ji} = \frac{P}{\sin (\theta_i^* - \theta_{i+1}^* )}.
\label{eq:Coupling}
\end{equation}

\begin{figure}
\begin{center}
\includegraphics[width=0.5\textwidth]{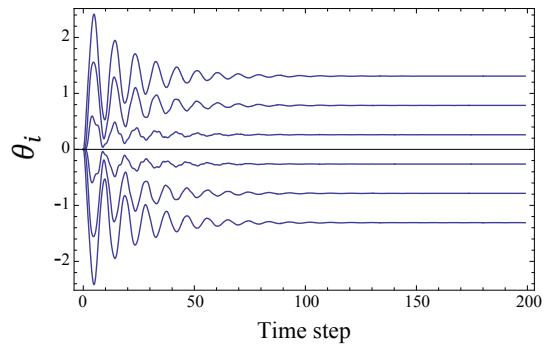}
\caption{Stationary Solution}
\label{fig:Stationary Solution}
\end{center}
\end{figure}

\begin{figure}
\begin{center}
\includegraphics[width=0.45\textwidth]{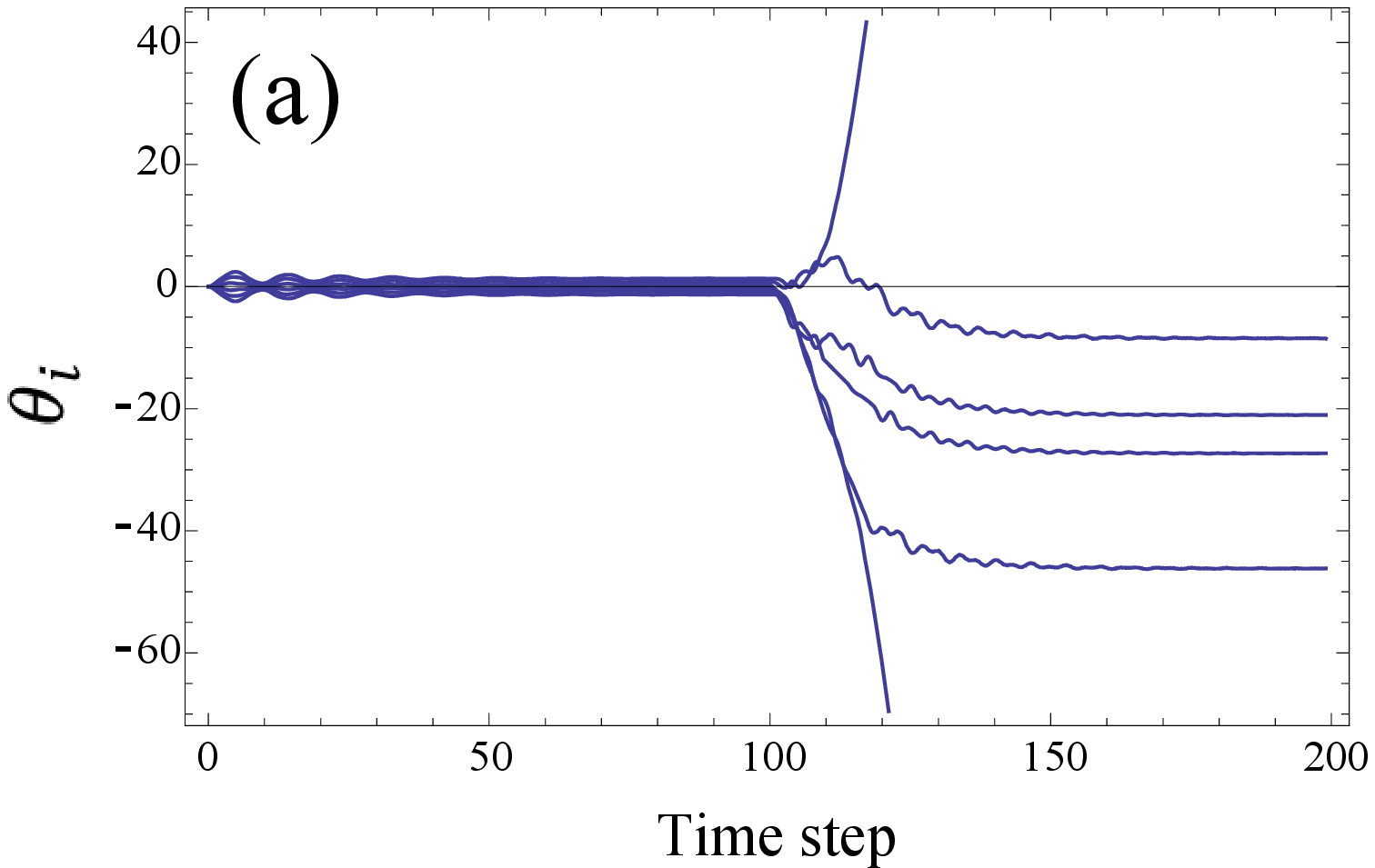}
\includegraphics[width=0.45\textwidth]{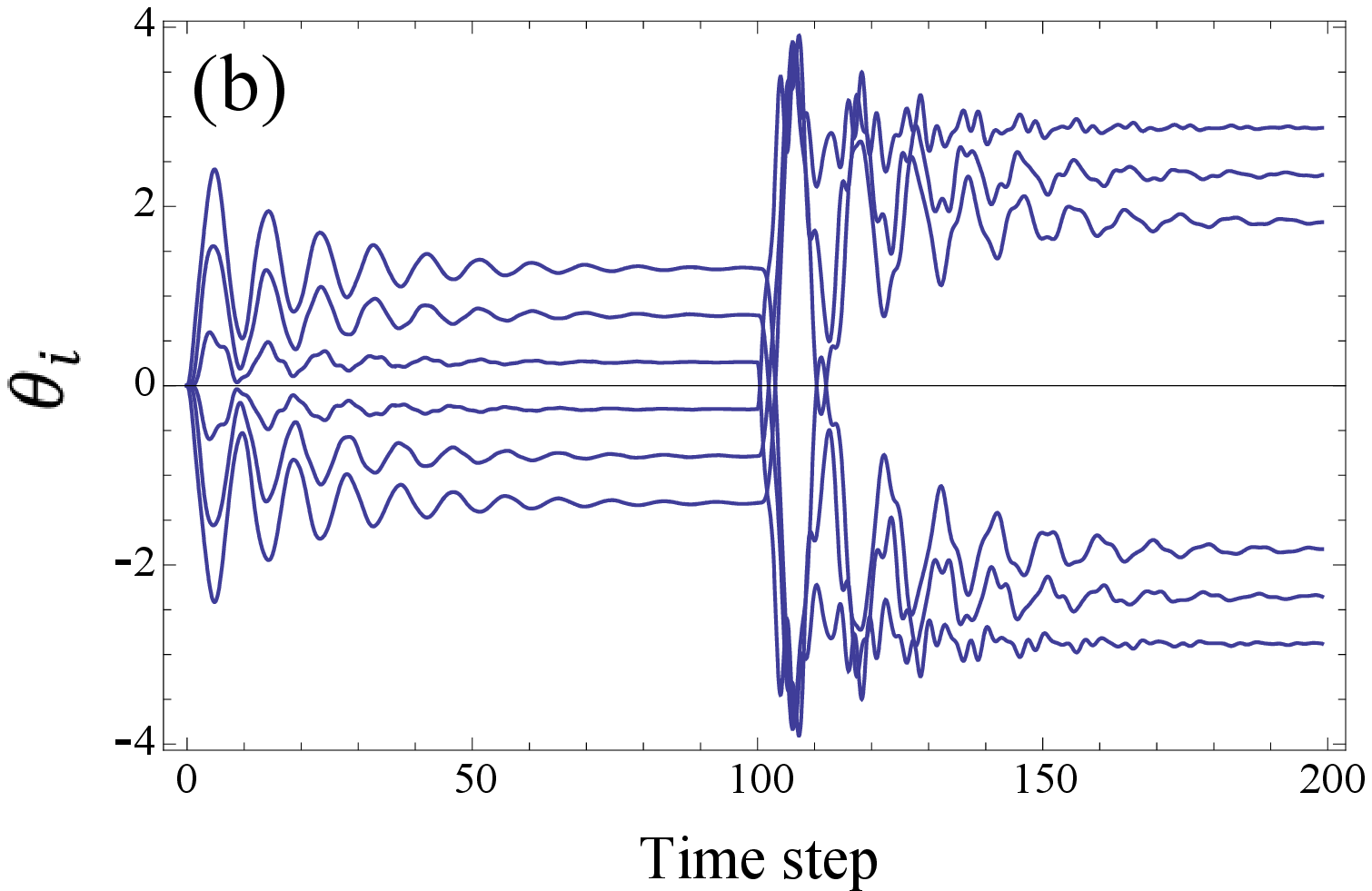}
\caption{Fluctuation and Elasticity}
\label{fig:Flucuation and Elastisity}
\end{center}
\end{figure}

We simulate the behavior of the system by solving Eq. (\ref{eq:Oscillator}) numerically with the initial condition $\dot{\theta}_i(0)=\ddot{\theta}_i(0)=0 (i=1, \cdots, N)$. 
Synchronization was reproduced as an equilibrium solution in a simple NN graph as shown in Fig. \ref{fig:Stationary Solution}. 
The response to a sectoral fluctuation at $t=100$ is also simulated in the NN graph.
In the case of zero elasticity ($\epsilon_d=0$), the synchronization was broken, as shown in Fig. \ref{fig:Flucuation and Elastisity} (a).
In contrast, in the case of finite elasticity ($\epsilon_d<0$), stability was restored after a shift of phase, as shown in Fig. \ref{fig:Flucuation and Elastisity} (b).

\section{Analysis of sectoral synchronization}

In this section, we analyze the sectoral synchronization observed in the Japanese business cycle using the coupled oscillator model described in the previous section.
The effects on synchronization resulting from a sectoral shock are studied for systems with different price elasticities and coupling strengths.

\subsection{Synchronization}

For the business cycle with a common period of $T=60$ months, the phases of production, shipment, and inventory are given in the article \cite{Iyetomi2011b}. 
The sectors are rearranged in decreasing order of the observed phase of production.
If we assume that the network is the NN graph, we can calibrated the coupling strengths $k_{ij}$ using Eq. (\ref{eq:Coupling}) with the observed phases $\theta_i^*$.
Analytical solutions of the stationary phases are then obtained using Eq. (\ref{eq:StatSol}). 

\begin{figure}
\begin{center}
\includegraphics[width=0.43\textwidth]{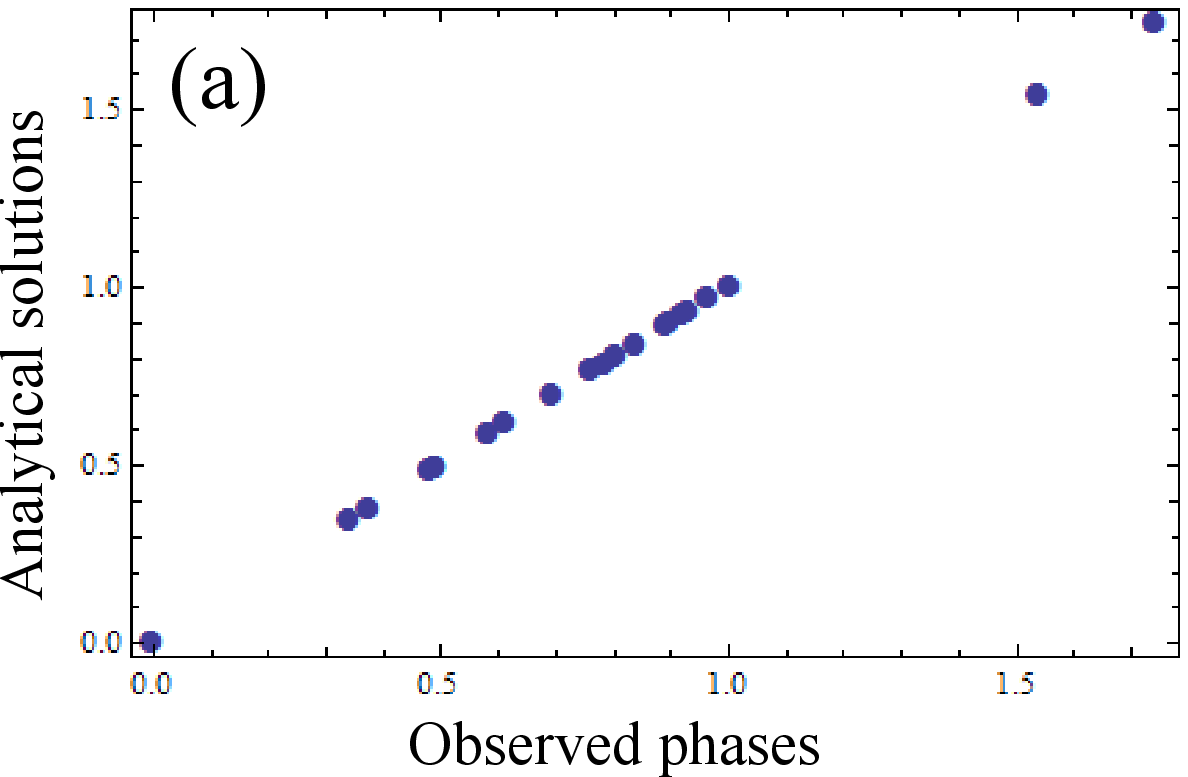}
\includegraphics[width=0.45\textwidth]{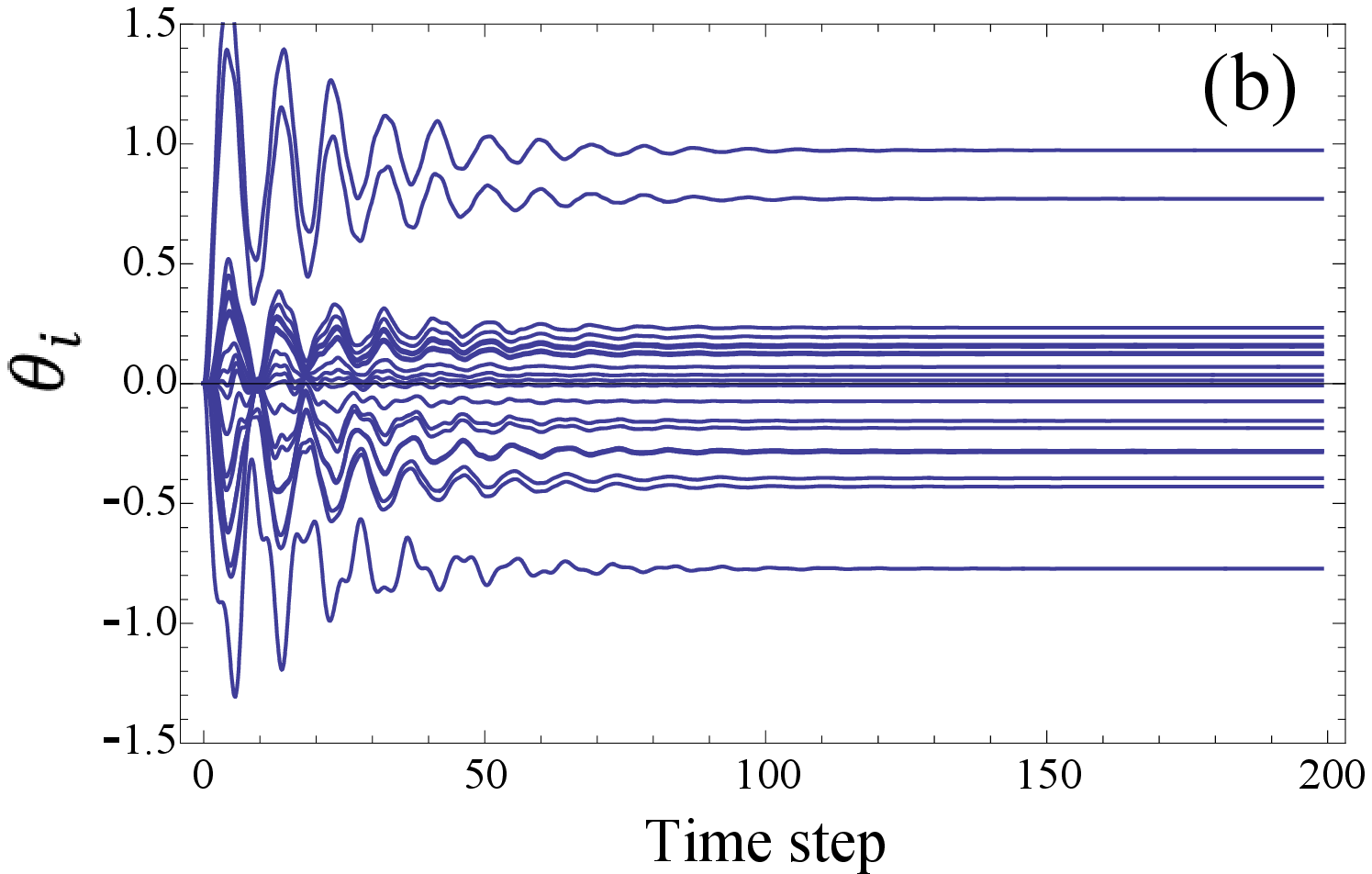}
\caption{Parameter calibration and synchronization}
\label{fig:IIPstationary}
\end{center}
\end{figure}

The analytical solutions of stationary phases are compared with the observed phases in Fig. \ref{fig:IIPstationary} (a).
The agreement between the analytical solutions and the observed phases is quite good.
We simulate the behavior of the system by solving Eqs. (\ref{eq:Oscillator}) numerically with the initial condition $\dot{\theta}_i(0)=\ddot{\theta}_i(0)=0 (i=1, \cdots, N)$. 
Synchronization was reproduced as an equilibrium solution in a simple NN graph, as shown in Fig. \ref{fig:IIPstationary} (b). 

\subsection{Fluctuating Goods Market}

The complex order parameter 
\begin{equation}
 q(t) = \frac{1}{N} \sum_{j=1}^N e^{i \theta_j (t)} = r ( cos(\phi) + i sin(\phi) )
\label{eq:OrderParam}
\end{equation}
is defined as a macroscopic quantity that corresponds to the centroid of the phases of oscillators \cite{Strogatz2000}.
The radius $r$ measures the coherence and $\phi$ is the average phase. 
If $\mathrm{Re}(q(t)) \approx 1$ and $\mathrm{Im}(q(t)) \approx 0$, the oscillators remain in the synchronization region where the phase differences are fairly small. 
In contrast, if $\mathrm{Re}(q(t))$ and $\mathrm{Im}(q(t))$ oscillate between $1$ and $-1$, the oscillators behave like a giant oscillator with a frequency additional to the common frequency $\omega$. 

\begin{figure}
\begin{center}
\includegraphics[width=0.45\textwidth]{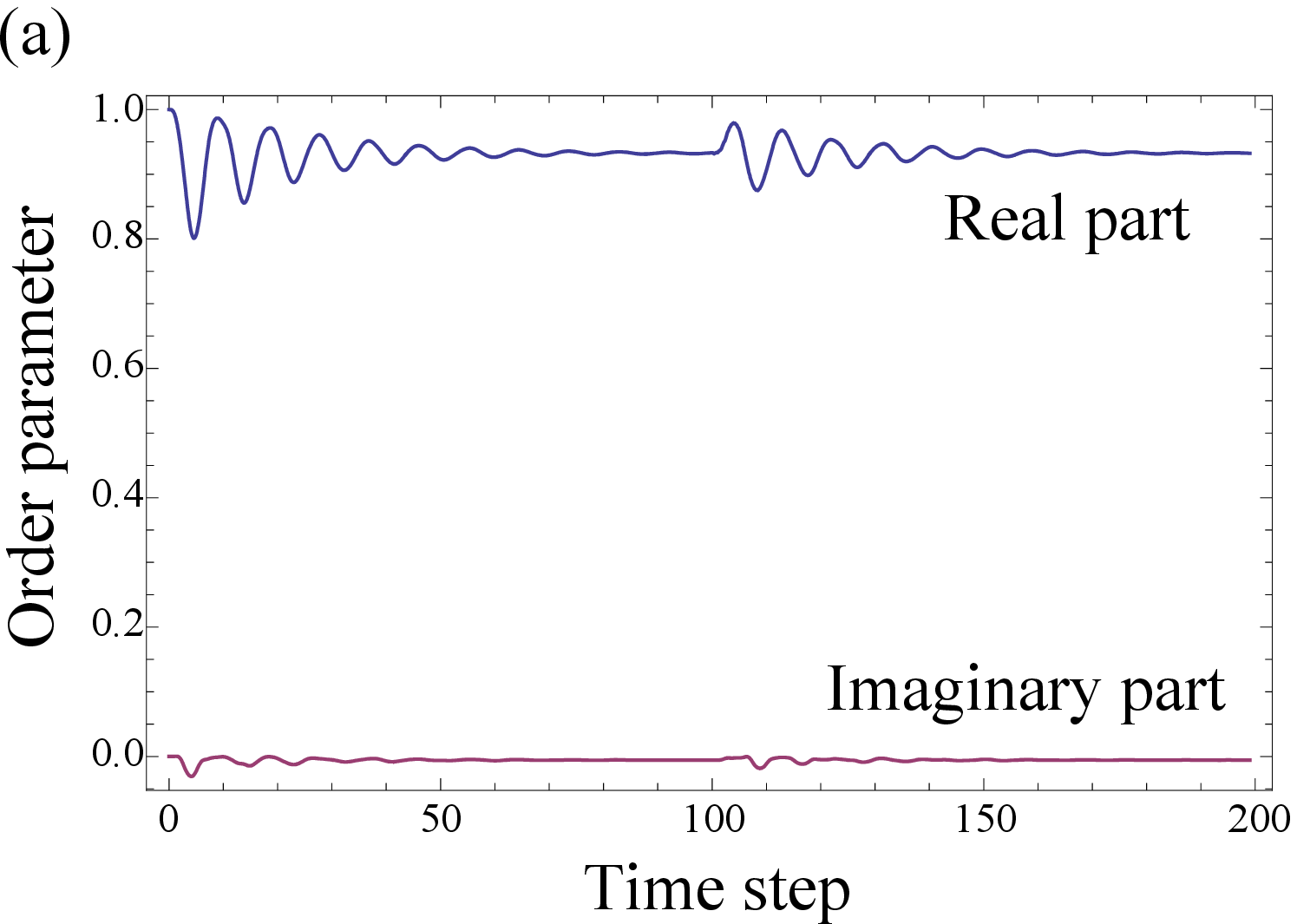}
\includegraphics[width=0.45\textwidth]{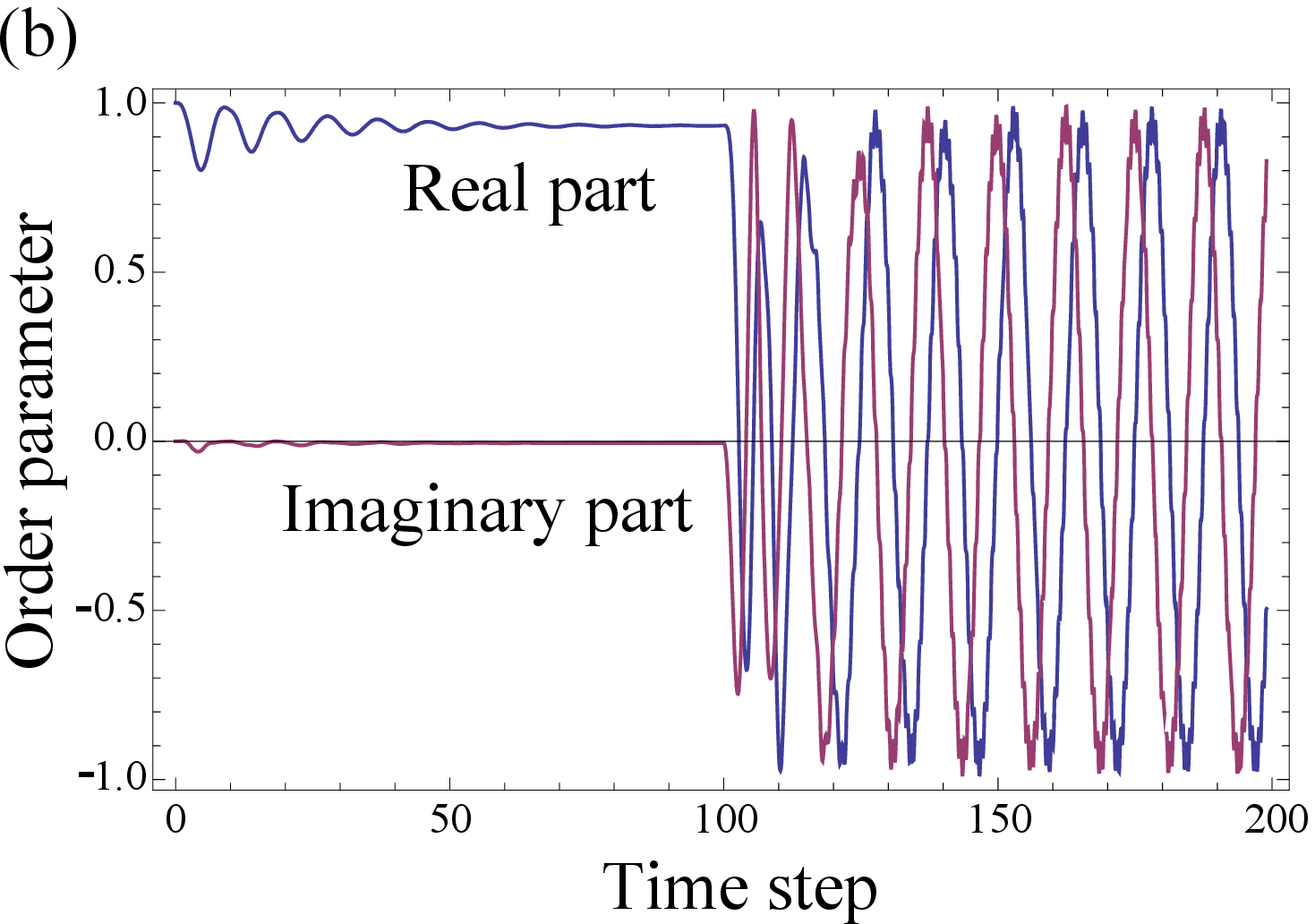}
\caption{Price elasticity  and stability of synchronization}
\label{fig:Flexibility}
\end{center}
\end{figure}

First, we simulate the response to the sectoral fluctuation at $t=100$ in the NN graph.
The result for a finite elasticity ($\epsilon_d<0$) is shown in Fig. \ref{fig:Flexibility} (a).
It is seen that $\mathrm{Re}(q(t)) \approx 1$ and $\mathrm{Im}(q(t)) \approx 0$, even after the sectoral shock was applied in the middle of the network at $t=100$.
This means that the synchronization is stable, i.e., the oscillators remain in the region where phase differences are fairly small.
For the case of zero elasticity ($\epsilon_d=0$), $\mathrm{Re}(q(t))$ and $\mathrm{Im}(q(t))$ oscillate rapidly between $1$ and $-1$ as shown in Fig. \ref{fig:Flexibility} (b).
This means that synchronization is broken and the oscillators behave like a giant oscillator with a high frequency additional to the common frequency $\omega$. 

\begin{figure}
\begin{center}
\includegraphics[width=0.45\textwidth]{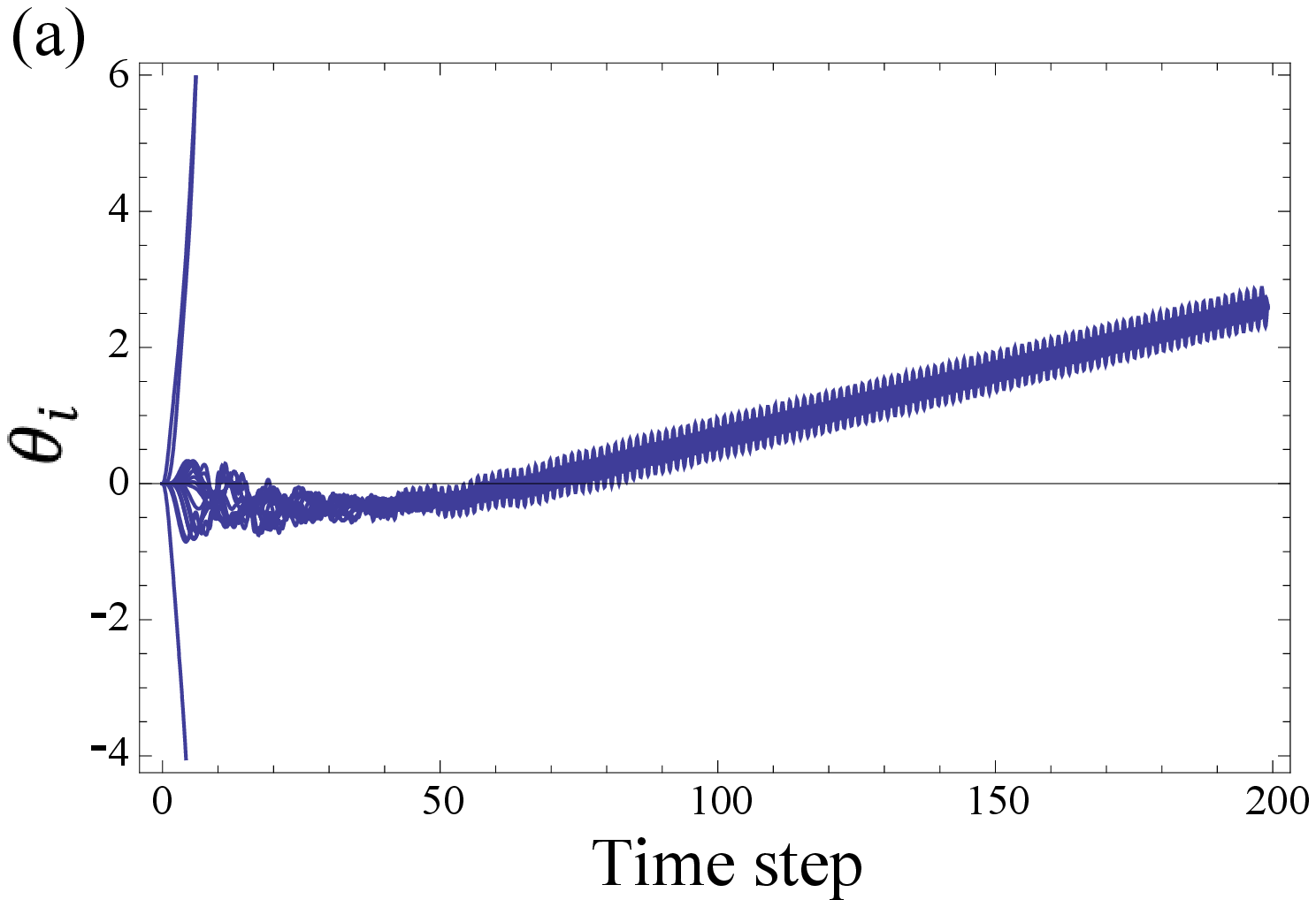}
\includegraphics[width=0.45\textwidth]{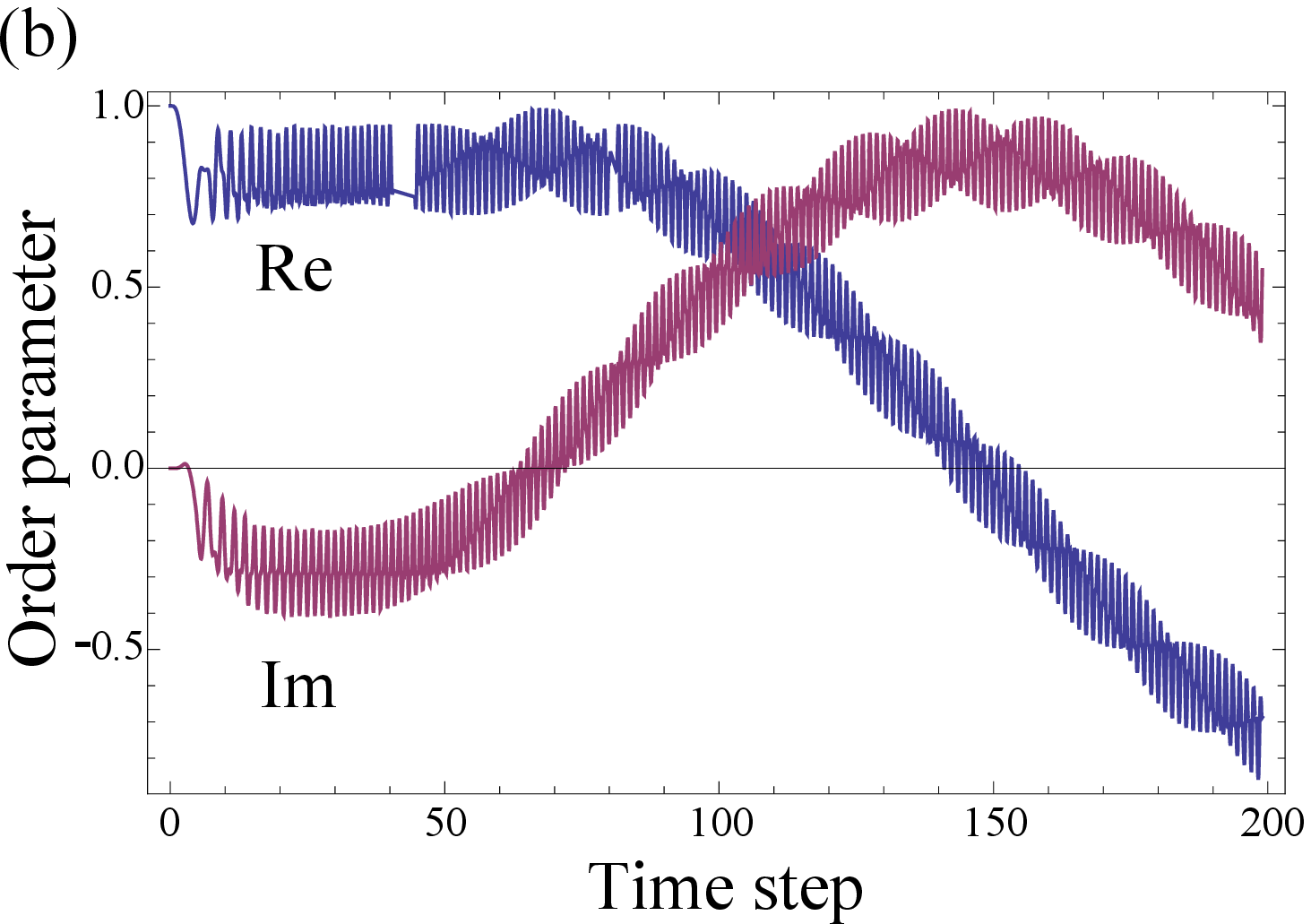}
\includegraphics[width=0.35\textwidth]{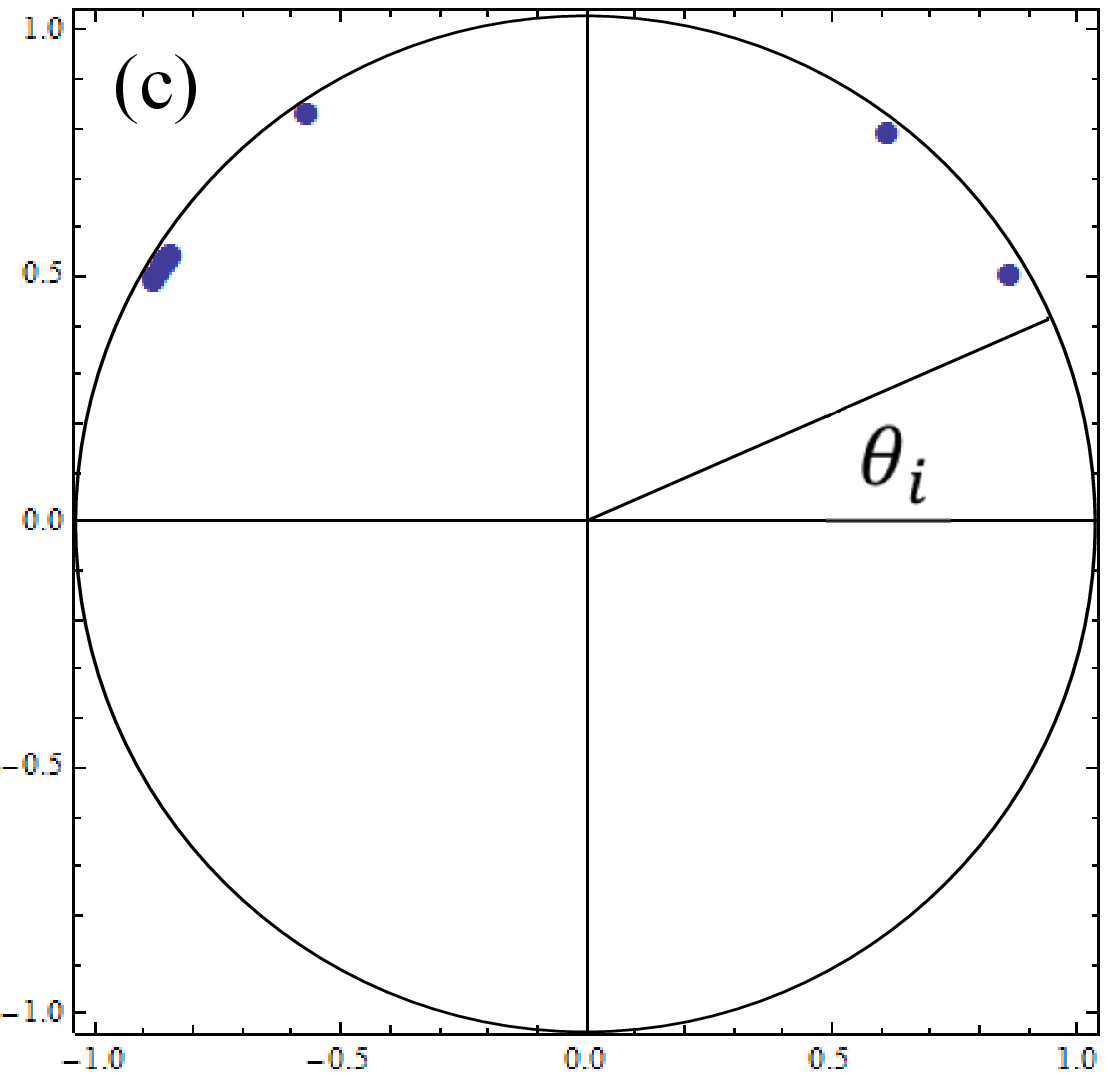}
\caption{Case with the coupling strength multiplied by 0.3}
\label{fig:kfact_0.3}
\end{center}
\end{figure}

Next, we simulate the oscillator system with weaker coupling strengths $k_{ij}$. 
We expect that the system to behave as if the oscillators were uncoupled for $k_{ij}$ below than a certain threshold $k_{ij}^c$.  
For the Kuramoto oscillator, the exact formula of the critical coupling strength $k_{ij}^c$ has been derived and verified with the results of numerical simulations.
The results of our simulation with weaker coupling strengths of $0.3 k_{ij}$ are shown in Fig. \ref{fig:kfact_0.3}.
The time evolution of the phases $\theta_i$ shown in Fig. \ref{fig:kfact_0.3} (a) depicts a few oscillators separating from the main part of the coupled oscillator system, which still exhibits synchronization.
The oscillator system acts like a giant oscillator with a low frequency additional to the common frequency $\omega$, as seen in Fig. \ref{fig:kfact_0.3} (b).
The phases $\theta_i$ at $t=200$ in Fig. \ref{fig:kfact_0.3} (c) show that the system disintegrated into four parts and that the main part moves with small phase differences. 

This result is quite different from our expectation, but this is also reasonable, 
because the system under consideration has $P_1=1, P_i=0 (i=2, \cdots, N-1), P_N=-1 (N=21)$.
Hence, one end of the system is pulled in the positive direction and the other end is pulled in the opposite direction.  
Therefore, if the coupling strengths are weak enough, the oscillators in both ends of the system are separated from the main part.
The oscillators in the main part lose tension and shrink to the small phase difference.  
However, it is noted that the system in the C graph with $P_i \neq 0 (i=1, \cdots, N)$ might behave similarly to the Kuramoto oscillator.


\section{Conclusions}

The sectoral synchronization observed for the Japanese business cycle in the Indices of Industrial Production data is an example of synchronization.
The stability of this synchronization under a shock, e.g., fluctuation of supply or demand, is a matter of interest in physics and economics.

First, existing theories for synchronization for economic systems, power systems, and physical and biological systems were reviewed. 

We then considered an economic system consisting of industry sectors and goods markets.
A coupled oscillator model exhibiting synchronization was developed based on the Kuramoto model with inertia by adding goods markets, and
analytic solutions of the stationary state and the coupling strength were obtained.

Finally, we analyzed the sectoral synchronization observed in the Japanese business cycle using the coupled oscillator model.
Effects on synchronization resulting from a sectoral shock were simulated for systems with different price elasticities and coupling strengths.
The agreement between the analytical solutions and the observed phases was quite good.
Synchronization was reproduced as an equilibrium solution in a simple NN graph. 
Analysis of the order parameters showed that the synchronization was stable for a finite elasticity ($\epsilon_d<0$), 
whereas the synchronization was broken and the oscillators behaved like a giant oscillator with a frequency additional to the common frequency $\omega$ for zero elasticity ($\epsilon_d=0$). 

In the future work, we intend to study
\begin{enumerate}
\item Simulation of the oscillator system for the complete graph with $P_i \neq 0 (i=1, \cdots, N)$.
\item Theory of phase transition for synchronization, including derivation of the exact formula for the critical coupling strength $k_{ij}^c$.
\item Development of the economic implications for sectoral fluctuations and the propagation of risk through the economic network.
\end{enumerate}

\section*{Acknowledgements}

The authors thank the Yukawa Institute for Theoretical Physics at Kyoto University. 
Discussions during the YITP workshop YITP-W-11-04 on ``Econophysics 2011 -The Hitchhiker's Guide to the Economy-'' were useful for completing this work.


\begin{thebibliography}{99}
\bibitem{Iyetomi2011a} 
H.~Iyetomi, Y.~Nakayama, H.~Yoshikawa, H.~Aoyama, Y.~Fujiwara, Y.~Ikeda, and W.~Souma ``What Causes Business Cycles? - Analysis of the Japanese Industrial Production Data", Journal of the Japanese and International Economies 25, 246-272 (2011).
\bibitem{Iyetomi2011b} 
H.~Iyetomi, Y.~Nakayama, H.~Aoyama, Y.~Fujiwara, Y.~Ikeda, and W.~Souma, ``Fluctuation-dissipation theory of input-output inter-industrial relations", Phys Rev E 83, 016103 (2011)
\bibitem{Goodwin1947}
R.~M.~Goodwin, ``Dynamical coupling with especial reference to markets having production lags", Econometrica 15, 181-204 (1947).
\bibitem{Goodwin1951}
R.~M.~Goodwin, ``The nonlinear accelerator and the persistence of business cycles", Econometrica 19, 1-17 (1951).
\bibitem{LongPlosser1983} 
J~.B.~Long  and C.~I.~Plosser, ``Real business cycles", Journal of Political Economy 91, 39-69 (1983).
\bibitem{LongPlosser1987} 
J~.B.~Long  and C.~I.~Plosser, ``Sectoral and aggregate shocks in the business cycle", American Economic Review 77, 333-336 (1987).
\bibitem{Anderson1999}
H.~M.~Anderson and J.~B.~Ramsey, ``U.S. and Canadian Industrial Production Indices As Coupled Oscillators", Economic research reports PR \# 99-01, New York University (1999).
\bibitem{Selover2003}
D.~D.~Selover, R.~V.~Jensen, and J.~Kroll, ``Industrial Sector Mode-Locking and Business Cycle Formation", Studies in Nonlinear Dynamics \& Econometrics7, 1-37 (2003).
\bibitem{Sussmuth2003}
B. Sussmuth, ``Business Cycles in the Contemporary World", Springer Berlin Heidelberg (2003).
\bibitem{Kundur1993}
P.~Kunder, ``Power System Stability and Control", 221-225 McGraw-Hill New York (1993).
\bibitem{Kuramoto1975} 
Y.~Kuramoto, 1975, in International Symposium on Mathematical Problems in Theoretical Physics, edited by H. Araki, Lecture Notes in Physics No. 30 sSpringer, New Yorkd, p. 420.
\bibitem{Strogatz2000} 
S.~H.~Strogatz, ``From Kuramoto to Crawford: exploring the onset of synchronization in populations of coupled oscillators'', Physica D 143, 1-20 (2000). 
\bibitem{Acebron2005} 
J.~A.~Acebron, L.~L.~Bonilla, C.~J.~P.~Vicente, F.~Ritort, and R.~Spigler, ``The Kuramoto model: A simple paradigm for synchronization phenomena'', Rev Mode Phys 77, 137-185 (2005).
\bibitem{Filatrella2008} 
G.~Filatrella, A.~H.~Nielsen and N.~F.~Pedersen, ``Analysis of a power grid using a Kuramoto-like model'', European Physics Journal B, 61, 485-491 (2008).


\end{thebibliography}
\end{document}